\documentstyle[aps,twocolumn,floats,epsf]{revtex} 
\begin{document}  
\draft
\title{Quantum Chaos, Irreversible Classical Dynamics and Random Matrix 
Theory}                               
\author{A.~V.~Andreev$^{a,b}$, O.~Agam$^a$, B.~D.~Simons$^c$, 
and B.~L.~Altshuler.$^{a,b}$}  
\address{$^a$NEC Research Institute, 4 Independence Way, Princeton,
 NJ 08540, USA \\
$^b$Department of Physics, Massachusetts Institute of Technology, 77  
Massachusetts Avenue, Cambridge, MA\ 02139, USA\\
$^c$Cavendish Laboratory, Madingley Road, Cambridge, CB3\ 0HE, UK}
\maketitle     
\begin{abstract}
The Bohigas--Giannoni--Schmit conjecture stating that 
the statistical spectral properties of systems which are 
chaotic in their classical limit 
coincide with random matrix theory is proved. 
For this purpose a new semiclassical field theory 
for individual chaotic systems is constructed in
the framework of the non--linear $\sigma$-model. The low lying modes 
are shown to be associated with the Perron--Frobenius spectrum of 
the underlying irreversible classical dynamics. It is shown that 
the existence of a gap in the Perron-Frobenius spectrum 
results in a RMT behavior. Moreover, our formalism offers a way of 
calculating system 
specific corrections beyond RMT. 
\end{abstract}     
\par                                                                 

\vspace{0.5cm}  
The theory of random matrices~\cite{Mehta91} emerged from the need to
characterize complex quantum systems in which knowledge of the 
Hamiltonian is minimal, e.g. complex nuclei. The basic hypothesis
is that the Hamiltonian may be treated as one drawn from an ensemble of
random matrices with appropriate symmetries. It has been
proposed by invoking the complexity of systems which have 
many degrees of freedom with unknown interaction coupling among them. 

The study of the statistical quantum properties of systems with 
{\em small number} of degrees of freedom, within the framework of
random matrix theory (RMT), has developed along two parallel lines. 
The first was by considering an ensemble of random systems 
such as disordered metallic grains \cite{Altshuler85}. 
Randomness in this case is introduced on the level of the Hamiltonian 
itself, e.g. as a consequence of the unknown impurity configuration.
In the second approach, RMT was used in order to understand the level 
statistics of non-stochastic systems which are chaotic in their classical 
limit such as the Sinai or the stadium billiards \cite{Bohigas84}.
Here ``randomness'' is generated by the underlying deterministic
classical dynamics itself. Nevertheless, it has been conjectured 
\cite{Bohigas84} that ``spectrum fluctuations of quantal 
time--reversal invariant systems whose classical analogues are strongly 
chaotic have the  Gaussian Orthogonal Ensembles pattern''.

Despite being supported by extensive numerical studies, the origin of the 
success of RMT as well as its domain of validity are still not completely 
resolved. In this letter we show that, in the semiclassical limit, this
conjecture is indeed valid for systems with exponential decay of 
classical correlation functions in time. Moreover, the formalism 
which we introduce below offers a way of calculating system specific 
corrections beyond RMT. 

So far, the main attempts to establish the 
relationship between non--stochastic chaotic systems and RMT, 
have been based on {\em periodic orbit theory} \cite{Gutzwiller90}.
Gutzwiller's trace formula expresses the semiclassical density of 
states as an infinite sum over the classical periodic orbits of 
the system. However, the number of periodic orbits is exponentially 
large and clearly contains information that is redundant from 
quantum mechanical point of view. This detailed information 
conceals the way of drawing a connection between the quantum behavior 
of chaotic systems and RMT. Indeed,  the success of the periodic orbit theory
approach in reproducing RMT results~\cite{Berry85} appears to be limited.

Here we develop a new semiclassical approach in which the basic
classical ingredients are not individual periodic orbits but global 
modes of the time evolution of the underlying classical system. 
It is possible to construct a field theory in which the effective 
action is associated with the classical flow in phase space. We argue 
that, the statistical quantum properties of the system 
are intimately related to the {\em irreversible} 
classical chaotic dynamics or, more precisely, to the 
Perron--Frobenius  modes, in which a disturbance in the classical 
probability density of chaotic system relaxes into the ergodic 
distribution. These modes decay at different rates. This, in principle, 
enables a description of the system at levels of increasing complexity by 
incorporating higher and higher modes. The ``zero mode'' manifests
the conservation of classical probability and corresponds to a 
uniform distribution over the energy shell. We show that RMT 
coincides with a description which takes into account only this mode.
The deviations from the universal RMT behavior emerge from the higher 
modes of the evolution of the classical system. 

Our approach is analogous to that of disordered systems 
where the diffusion modes account for the classical relaxation.
However, in the field theoretic description of disordered 
systems \cite{Efetov83} averaging is performed over 
an ensemble. Here, in contrast, in order to characterize 
{\em individual} systems only energy averaging will be employed. 

The main result of this letter will be to establish the Bohigas-Giannoni-Schmit
conjecture. To do so we will first show that quantum statistical correlators
are described by a functional non-linear $\sigma$-model. Its low lying modes 
are identified with the Perron-Frobenius eigenmodes of the 
underlying classical dynamics. Then we argue that, provided classical 
correlation functions decay exponentially in time, there is an
energy domain where the zero mode contribution governs the behavior.
This follows from the fact that in such systems the Perron-Frobenius 
spectrum has a gap. Finally we establish 
the relation to RMT by identifying it with the zero mode of the
constructed field theory.

Let $\hat{H}$ define a quantum Hamiltonian whose classical counterpart,
$H({\bf x})$, is chaotic. Here we use the notation 
${\bf x}=({\bf q},{\bf p})$ to denote a vector defined in a $2\times d$ 
dimensional phase space. We restrict
attention to a system in which all classical trajectories are 
unstable, and there are no islands of regular motion in phase space. 
At sufficiently high energy $E_0$, the mean spacing
$\Delta$ between adjacent energy levels can be assumed constant and 
expressed through the Weyl formula,
\begin{equation} 
\frac{1}{\Delta} =\frac{1}{(2\pi \hbar)^d}\int d^{2d} x \ 
\delta\left[E- H({\bf x})\right]. 
\label{weyl} 
\end{equation} 
Henceforth energy and time will be measured in units 
of $\Delta$ and the Heisenberg time $\hbar/\Delta$ respectively. 

The energy averaging is taken over an energy band containing a large 
number of levels $N$ 
such that $1\ll N \ll \epsilon_0 = E_0/\Delta$. For 
simplicity we choose to work with Gaussian averages $(\epsilon=E/\Delta)$:
\begin{equation} 
\langle \cdots \rangle_{\epsilon_0}=\int  
\frac{d\epsilon}{\sqrt{2\pi} N } \exp\left[ 
-\frac{(\epsilon-\epsilon_0)^2}{2N^2}\right] (\cdots ).
\label{av} 
\end{equation} 

The basic quantity calculated within the field theory approach is the 
generating function $\langle {\cal Z}(J)\rangle_{\epsilon_0}$. 
Any $n$-point correlation function can be obtained by taking its
derivatives with respect to the various components of the source $J$. 
However, to keep the discussion simple, we will restrict attention to systems
belonging to the unitary ensemble (i.e. those with broken T-invariance),
and focus on two-point correlation functions such as the correlator of 
density of states, $R(s)=\langle \rho(\epsilon+s) \rho(\epsilon) 
\rangle_{\epsilon_0}-1$, where $\rho(\epsilon)={\rm Tr} \delta 
(\epsilon- H)$. As long as $s$ is sufficiently small 
compared to the bandwidth $N$, the final results
are independent of the particular form of energy averaging.

Following the usual approach~\cite{Efetov83}, the generating function can
be expressed as a field integral. Introducing the four component 
supervector field $\Psi^T({\bf q})=
(\psi^R,\chi^R,\psi^A \chi^A)$, where $\psi$ and $\chi$ denote 
commuting and anti-commuting components respectively, and the superscript
$A$ ($R$) denotes the advanced (retarded) components \cite{Efetov83}, 
we can define
\begin{equation} 
{\cal Z}(J) = \int {\cal D}\Psi\ e^{-{\cal S}[\Psi,\epsilon]},
\label{Gfunction}
\end{equation} 
where the action is given by 
\begin{equation} 
{\cal S} [\Psi,\epsilon] = i \int \! d^dq 
\bar{\Psi}({\bf q}) \left[\epsilon-\frac{s^+}{2}\Lambda
 - \hat{H} - Jk\Lambda \right]\Psi({\bf q}). 
\label{Zaction}
\end{equation} 
Here $s^+=s+i0$, while $\Lambda=\mbox{diag}(1,1,-1,-1)$ and 
$k=\mbox{diag}(1,-1,1,-1)$  break the symmetry between 
retarded/advanced field components and supersymmetry respectively,
and $\bar{\Psi}=\Psi^\dagger \Lambda$.
The use of fermionic as 
well as bosonic components ensures the normalization 
${\cal Z}(0)=1$. To evaluate the two point
density correlator one can choose $J$ to be constant, then 
\begin{equation}
R(s)= -\frac{1}{16 \pi^2 } \Re \left. \frac{\partial^2 
\langle{\cal Z}(J)\rangle_{\epsilon_0}}{\partial J^2} \right|_{J=0}. 
\end{equation}

Energy averaging (\ref{av}) of ${\cal Z}(J)$ generates
a quartic interaction of the $\Psi$ fields: 
${\cal S}(\Psi, \epsilon) \to {\cal S}(\Psi,\epsilon_0)
+ {\cal S}_{\mbox{i}}$, where
\begin{equation} 
{\cal S}_{\mbox{i}}= \frac{ N^2}{2} \left( \int d^dq 
\bar{\Psi}({\bf q})\Psi({\bf q}) \right)^2.
\end{equation}
In contrast to impurity averaging~\cite{Efetov83}, energy averaging
induces a {\em non--local} interaction of the $\Psi$ fields. This
interaction term can be decoupled by means of Hubbard--Stratonovich
transformation, with the introduction of $4\times 4$ supermatrix fields
$Q({\bf q},{\bf q'})$ which depend on two coordinates,
\begin{equation} 
\label{newhubstrat} 
\exp \left( -{\cal S}_{\mbox{i}} \right) =\int \! {\cal D} Q  
\exp\left[- \mbox{STr}_{{\bf q}} \left(\frac{Q^2}{2}-i{\cal N}  
Q \Psi\bar{\Psi} \right)\right].
\end{equation} 
Here ${\rm STr}_{{\bf q}}$ denotes the trace operation for supermatrices, 
while the subscript ${\bf q}$ implies a further extension of 
the trace to include integration over all spatial variables, e.g.
${\rm STr}_{{\bf q}}Q^2=\int d^dq d^d q' {\rm STr} Q({\bf q},{\bf q'})
Q({\bf q'},{\bf q})$.   Integrating over $\Psi$ we obtain:
\begin{eqnarray} 
\label{genfunc} 
\langle {\cal Z}(J)\rangle_{\epsilon_0}=
\int {\cal D} Q\ \exp\left[{-\frac{1}{2} \mbox{STr}_{{\bf q}} Q^2+ 
\mbox{STr}_{{\bf q}} \ln {{\cal G}}}\right], \\
\label{efftheory} 
{{\cal G}}^{-1}(Q)= \epsilon_0-\frac{s^+}{2}\Lambda - \hat{H}
-Jk\Lambda-N Q. 
\end{eqnarray}
The derivation of Eq.~(\ref{genfunc}) involved no approximation. Further
progress is possible only within a saddle-point approximation, 
which is accurate to order $1/N$. Varying the total action with 
respect to $Q$ one obtains the saddle-point equation,
\begin{equation}
Q_0 {{\cal G}}^{-1}(Q_0) = N\label{suddle}
\end{equation}
where $Q_0$ and ${{\cal G}}(Q_0)$ are operators. 

To understand the structure of the saddle-point manifold, in
the semiclassical limit, it is useful to employ the Wigner representation
of operators. Given an operator $\hat{{\cal O}}$ as a set of matrix elements
${\cal O}({\bf q}_1,{\bf q}_2)$  between two position states at ${\bf q}_1$ 
and ${\bf q}_2$, its Wigner representation is a function of the phase 
space variables ${\bf x}$ defined by ${\cal O}({\bf x}) = \int 
d^dq' \exp (i {\bf p}{\bf q'}/\hbar) {\cal O}({\bf q}+{\bf q'}/2,
{\bf q}-{\bf q'}/2)$. We will use the fact that, 
in the semiclassical limit, the Wigner transform of a product of operators
is equal the product of the Wigner transformed operators, namely, 
$({\cal O}_1{\cal O}_2) ({\bf x})\to {\cal O}_1({\bf x})
{\cal O}_2 ({\bf x})$, where ${\cal O}_{1,2}({\bf x})$ are 
smooth slowly varying functions on quantum scale. 

Treating $s$ and $J$ as small compared to the bandwidth $N$, and 
introducing the phase space variables, ${\bf x}_\parallel$ on the 
energy shell and $x_\perp=H$ perpendicular to the energy shell, the 
solution of Eq.~(\ref{suddle}) can be written in the Wigner 
representation as
\begin{equation}
Q_0 ({\bf x}) =\frac{ \epsilon_0-H }{2N}
+ i\left[ 1 - \left( \frac{\epsilon_0-H}{2N}\right)^2
\right]^{\frac{1}{2}} \Lambda . 
\end{equation} 
This solution is, however, not unique. In fact, the saddle-point solutions 
form a degenerate manifold in superspace associated with the group of 
pseudo-unitary rotations. Both the integration measure and the action in 
Eq.~(\ref{genfunc}) are invariant under the group of transformations
$Q({\bf x}) \to U^{-1}(x_\perp)Q ({\bf x})U(x_\perp)$, where 
$U(x_\perp)$ belongs to the pseudo-unitary super-group  
$U(1,1/2)$~\cite{Zirnbauer86}. Thus, any matrix of the form $Q(x_\perp) = 
U^{-1}(x_\perp)Q_0 (x_\perp)U(x_\perp)$ is a solution of the saddle-point 
equation (\ref{genfunc}).

When integrating over the fluctuations $\delta Q$, near the saddle-point 
manifold, one has to take into account the anisotropy of the dependence of the
action on the fluctuations $\delta Q$. Those fluctuations on which the action
depends strongly (massive modes) can be integrated out within a conventional
saddle-point approximation. The remaining fluctuations describe the Goldstone 
modes of the system, and their integration must be performed exactly. 
The experience gained from studies of disordered systems 
suggests that these degrees of freedom
can be parametrized by $Q({\bf x}) = T^{-1}({\bf x})Q_0(x_\perp) T({\bf x})$. 
However, integration over the massive modes \cite{big-paper} shows that, 
in the limit $N\gg 1$,
the only non-vanishing contribution comes from matrices $T$ which are 
independent of the energy $x_{\perp}$. The Goldstone modes can be therefore 
parametrized by 
\begin{equation}
Q({\bf x}) = T^{-1}({\bf x}_\parallel)Q_0(x_\perp)
T({\bf x}_\parallel), \label{TQT}
\end{equation}
where $T(x_\parallel)$ belong to the coset space 
$U(1,1/2)/U(1/1)\otimes U(1/1)$ \cite{Zirnbauer86}.

The derivation of the effective field theory can thus be obtained by:
(i) Substituting Eq.~(\ref{TQT}) into Eq.~(\ref{genfunc}), approximating 
commutators by Poisson brackets $[{\cal O}_1,{\cal O}_2] 
({\bf x}) \to i\hbar \{{\cal O}_1({\bf x}),{\cal O}_2({\bf x})\}$,
and replacing the trace by the phase space integral, 
${\rm Tr}_{\bf q} ({\cal O})
=h^{-d} \int d{\bf x}{\cal O}({\bf x})$ (this is the entry point
of the semiclassical analysis) (ii) Expanding the logarithm to 
first order in $s^+$, $J$, and the Poisson bracket 
$\left\{H,T({\bf x}_\parallel)\right\}$ (higher order terms of this expansion 
appear at order $1/N$ and are neglected) (iii) Performing the $x_\perp$ 
integration of the resulting action. The last step relies on the fact that 
within the energy band, where averaging takes place, the classical 
dynamics is independent of the energy. As a result we obtain the 
$\sigma$-model: 
\begin{eqnarray}
\langle {\cal Z}(J) \rangle_{\epsilon_0} = 
\int{\cal D}T({\bf x}_\parallel) \exp \left( -{\cal S}_{\mbox{eff}} \right),
\label{sigma-model}\\
{\cal S}_{\mbox{eff}}= \frac{i \pi}{h^d}  \!\int \! d{\bf x}_{\parallel}
 \mbox{STr}
\left[\left( \frac{s^+}{2} \Lambda +Jk\Lambda\right){\cal Q}- i{\cal Q}T^{-1}
{\cal L}T \right] \label{effective-action},
\end{eqnarray}
where
\begin{equation}
{\cal Q}=-\frac{i}{\pi N}\int dx_\perp T^{-1}({\bf x}_\parallel) Q_0(x_\perp)
T({\bf x}_\parallel)=T^{-1}\Lambda T,
\end{equation}  
and ${\cal L}$ is the dimensionless infinitesimal time evolution operator 
defined by the Poisson bracket 
\begin{equation}
{\cal L} \ \cdot=\hbar \  \{ H, \ \cdot \ \}. \label{Liouville}
\end{equation}

For a stochastic Hamiltonian, an action equivalent to 
Eq.~(\ref{effective-action}) has been proposed recently 
using an argument which relies on disorder averaging in
the limit of vanishing disorder \cite{Dima}. 

To make sense of the functional integral in Eq.~(\ref{sigma-model})
we must identify the low lying modes of the action. In the case of 
impurities, the low lying degrees of freedom correspond to the eigenvalues 
of the diffusion operator that form a discrete spectrum. In the present 
case it is tempting to associate the low energy degrees of freedom with 
eigenmodes of the unitary (reversible) evolution operator 
$e^{-{\cal L}t}$. However this identification is incorrect. 

As with any functional integral there is a need 
to define an appropriate regularization. For example the functional 
integral in Eq.~(\ref{sigma-model}) may be understood as 
the limit $a\to 0$ of a product of definite integrations 
over a discretized space, where 
$a$ denotes the discretization cell size. This admits to functions 
$T({\bf x}_\parallel)$ which are {\em smooth} and square integrable. 
More generally, a regularization can be obtained using a truncation 
of an arbitrary complete basis.   

In seeking such a basis, the eigenfunctions of the 
classical evolution operator ${\cal L}$ seem to be the 
natural choice. However, the intricate nature of chaotic 
classical evolution cause these eigenfunctions to lie generally 
outside the Hilbert space. This can be seen from the following 
consideration: Chaotic dynamics of probability densities
involves contraction along stable manifolds, together with 
stretching along unstable ones. Thus, in the course of time, an 
initially non-uniform distribution turns into a singular function on 
the unstable manifold, which in turn covers the whole energy shell 
densely. Therefore the eigenfunctions of ${\cal L}$, which 
require the infinite time limit, are not square integrable and their 
contribution to the functional integral can not be directly 
recovered by the discretization procedure involved in evaluating the 
functional integral. 

Thus, in choosing a convenient basis one has to take account of 
the regularization. Its primary effect is to truncate the contraction along 
the stable manifold and thereby render the classical evolution 
{\em irreversible}. It is the eigenfunctions of this regularized classical
evolution operator that serve as a suitable basis for the quantum mechanical 
correlator.

A natural way of introducing a regularization is to express the functional 
integral in terms of a basis of eigenfunctions of the evolution 
operator to which a small noise has been added such that a 
diffusion on the energy shell is allowed. The eigenfunctions of the resulting 
operator, which is no longer antihermitean, are rendered smooth along 
the stable manifold and square-integrable. 
When the strength of the noise is taken to zero --- the spectrum of 
the resulting operator, known in the literature as the Perron-Frobenius 
operator, reflects intrinsic irreversible properties of the 
{\em purely classical dynamics} \cite{Gaspard95}. Other approaches
which recover this spectrum involved a use of symbolic dynamics 
\cite{Ruelle}, course graining of the flow dynamics in phase space
\cite{Nicolis}, and analytic continuation methods \cite{Hasegawa92}.
We remark here that the physical spectrum of the classical operator
${\cal L}$ appears when it propagates smooth probability densities.
Thus the matrices $T({\bf x}_\parallel)$ over which ${\cal L}$ acts 
are understood to be smooth. In this respect, quantum mechanics 
can viewed as the natural framework for calculating irreversible 
properties of classically chaotic systems.

Let $\{ \gamma_n \}$ be the set of the eigenvalues of the 
generator, ${\cal L}$, of the Perron-Frobenius operator.
In ergodic systems, the leading eigenvalue $\gamma_0=0$ 
is non-degenerate, and manifests the conservation of probability density.
Thus any initial density relaxes, in the course of time, to the state 
associated with $\gamma_0$. If, in addition, this relaxation is
exponential in time, then the Perron-Frobenius spectrum 
has a gap associated with the slowest decay rate. Thus, 
for the first nonzero eigenvalue $\gamma_1$ we have 
$\gamma_1'\equiv {\rm Re}(\gamma_1)> 0$. This gap sets 
the ergodic time scale, $\tau_c=1/\gamma_1'$, 
over which the classical dynamics relaxes to equilibrium. 
In the case of disordered metallic grains, it coincides with the Thouless
time, while in ballistic systems or billiards it is of order of the time 
of flight across the system.
    
In the limit $s\ll \gamma_1'$ the dominant contribution
to Eq.~(\ref{sigma-model}) comes from the ergodic classical distribution,
the zero-mode ${\cal L}T_0=0$. With this contribution alone, 
the functional integral  (\ref{sigma-model}) becomes definite,
\begin{equation}
\langle {\cal Z}(J) \rangle_{\epsilon_0} = \int
d{\cal Q}_0  \exp \left( -i\frac{\pi}{2} \mbox{STr}
\left[\left( s^+ \Lambda +2Jk\Lambda\right){{\cal Q}}_0\right] \right),
\end{equation}
where ${{\cal Q}}_0= {T}_0^{-1} \Lambda {T}_0$. This expression coincides
with that for RMT and leads to the Wigner--Dyson statistics 
\cite{Efetov83,Verbaarschot85}.

This result can be generalized  to any $n$-point correlation 
function as well as to systems with T-invariance. 
One thus concludes that the quantum statistics of chaotic systems with 
exponential classical relaxation are described by RMT at energies smaller
than $\gamma_1'$.

RMT description is expected to hold even for certain chaotic
systems where the Perron-Frobenius spectrum is gapless~\cite{Bohigas84}, 
such as the 
stadium or the Sinai billiards where classical correlation functions 
decay algebraically in time \cite{Bunimovich}.
The resolvent $1/(z-{\cal L})$, in this case, is expected to have 
cuts which reach the $\Im z$ axis. Nevertheless, we expect the RMT 
description to hold whenever the spectral weight of the resolvent 
inside the strip $0\le \Re z \le 1$  (which excludes the pole 
at the origin, however) is much smaller than unity.

The strength of the field theoretic approach is that it encompasses a range 
of energy scales which go well beyond RMT. In particular, a similar
procedure to that used in Ref.~\cite{ab} can be employed to obtain 
$R(s)$ corresponding to the unitary ensemble
\begin{equation}
R(s)=-\frac{1}{4\pi^2}\frac{\partial^2}{\partial s^2} \ln {\cal D}(s) +
\frac{\cos (2 \pi s)}{2 \pi^2} {\cal D}(s),
\end{equation}
where ${\cal D}(s)=\prod_\mu A_\mu (s^2 + \gamma_\mu^2)^{-1}$ 
is the spectral determinant ($A_\mu$ being regularization factors~\cite{oba})
associated with the Perron-Frobenius spectrum. This 
confirms the conjecture made Ref.~\cite{oba}.

In conclusion, we have established the Bohigas--Giannoni--Schmit conjecture
for chaotic systems in which classical correlation functions decay 
sufficiently fast. Moreover, the field theoretic approach allows the study 
of statistical characteristics of such systems on an energy 
scale which is much wider than that in which RMT applies. These statistics 
are determined by the analytic properties of the classical resolvent
operator $1/(z-{\cal L})$. This theory, in principle, offers a 
systematic controlled way of investigating quantum corrections 
to the leading semiclassical description.  
       
We are grateful to D.~E.~Khmel'nitskii, B.~A. Muzykantskii, 
A.~M.~Polyakov, D.~Ruelle Ya.~G.~Sinai and N.~Taniguchi for 
stimulating discussions. This work was supported in part by 
JSEP No. DAAL 03-89-0001 and by NSF Grant No. No. DMR 92-14480.
O.~A.~acknowledges the support of the Rothschild Fellowship.

\end{document}